\documentclass[5p,twocolumn,times]{elsarticle}
\usepackage{dcolumn}
\usepackage[T1]{fontenc}
\usepackage{lmodern}
\usepackage[english]{babel}
\usepackage{amsmath,amsfonts,amsthm,amssymb,amsbsy, siunitx} \usepackage{mathtools, cuted}
\usepackage{braket}
\usepackage{bm}
\usepackage{booktabs}
\usepackage[labelfont=bf,justification=justified ,singlelinecheck=false,font=footnotesize]{caption}
\usepackage{lineno}
\modulolinenumbers[1]
\usepackage{adjustbox}
\usepackage[para,flushleft]{threeparttable}

\usepackage[final, authormarkuptext=name]{changes}
\definechangesauthor[name={A}, color=red]{A}
\usepackage{tablefootnote}
\usepackage{multirow}
\usepackage{url}
\usepackage{footnote}
\makesavenoteenv{tabular}
\makesavenoteenv{table}
\usepackage{gensymb}
\usepackage[version=4]{mhchem}

\begin{document}
	
\begin{frontmatter}
		
		\title{The Millimeter-Wave Spectrum of Doubly Deuterated Propylene Oxide $\mathrm{CH_{3}CHCD_{2}O}$}   
		
		\author[kassel]{Pascal Stahl\corref{mycorrespondingauthor}}
		\ead{p.stahl@physik.uni-kassel.de}
		\author[kasselc]{Denis Kargin}
		\author[kasselc]{Rudolf Pietschnig}
		\author[kassel]{Thomas F. Giesen}
		\author[kassel]{Guido W. Fuchs}
		
		\address[kassel]{Institute of Physics, University of Kassel,
              Heinrich-Plett-Str. 40, 34132 Kassel, Germany}
\address[kasselc]{Institute of Chemistry, University of Kassel,
              Heinrich-Plett-Str. 40, 34132 Kassel, Germany}              
		\cortext[mycorrespondingauthor]{Corresponding author}

\begin{abstract}
 Spectra of doubly deuterated propylene oxide,  $\mathrm{CH_{3}CHCD_{2}O}$,  were recorded in the millimeter-wave spectral region up to 330 GHz utilizing a {\it 2f} frequency modulated Terahertz spectrometer. Rotational and centrifugal
distortion constants and tunneling parameters for the description of the internal rotation of the methyl group were retrieved. The software XIAM was used to describe the A-E tunneling splitting with a frequency uncertainty of 82 kHz. Molecular parameters derived from our measurements are complemented by quantum chemical anharmonic frequency calculations on the B3LYP/aug-cc-pVTZ level of theory. In addition, the barrier height to internal rotation of the methyl group,$V_3$, was derived experimentally and found to slightly differ from the value reported for the main isotopologue. In view of astrophysical observations we provide accurate line lists to search for doubly deuterated propylene oxide in space. Furthermore, we deliver improved molecular parameters to further investigate the properties of a simple chiral molecule possessing internal large amplitude motions.  
\end{abstract}


\end{frontmatter}

\section{Introduction}
Propylene oxide, CH$_3$C$_2$H$_3$O (\textit{alias} methyloxirane or propene oxide),  is a stable chiral molecule whose microwave spectrum up to 40 GHz was first reported by Herschbach \& Swalen  in 1957  \citep{Swalen.1957,Herschbach.1958}. Its high vapor pressure at room temperature and its commercial availability in high enantiomeric purity make it a well-suited object for studying chiral properties of gas-phase molecules. Propylene oxide (PO) is a simple epoxide ring with one hydrogen atom substituted by a methyl group. The large amplitude internal torsion of the methyl group in a triple well potential causes rotational levels of PO to split into sub-levels of A and E symmetry.  From the analysis of A-E split rotational lines in the ground, first and second excited torsion states, Herschbach \& Swalen derived the potential barrier height of 895(5) cm$^{-1}$. In 2017 Mesko \textit{et al.} reported new ground state spectra of PO up to 1 THz which enabled improved molecular parameters and a more accurate potential barrier height \citep{Mesko.2017}. Recently, rotational spectra of PO in the first torsional excited state up to 1 THz were analyzed, which show large A-E splittings \citep{Stahl.2021}. These new measurements helped to derive more accurate tunneling parameters and molecular constants of the large amplitude motion in the vibrational excited state. Ground state spectra of $\mathrm{^{13}C}$-substituted isotopologues reported in 1977 by Crewell \textit{et al.} \citep{Creswell.1977}, and multiply deuterated and $\mathrm{^{18}O}$ substituted species were measured  by Imachi \& Kuczkowski \citep{Imachi.1983}, which allowed them to experimentally derive the equilibrium structure of propylene oxide. In these studies, only lines of A symmetry were considered, leaving the question of the influence of isotopes on the tunneling barrier height unresolved.\\
In 2016, the detection of propylene oxide as the first chiral molecule discovered in space attracted great astrophysical interest. The observation of rotational lines towards the north core (N) of the giant molecular cloud SgrB2 \citep{McGuire.2016}
 raised the fundamental question on how chiral molecules are formed in space. Possible chemical formation routes have been discussed 
 \citep{Ellinger.2020}, 
 which also include reactions in or on cryogenic solids induced by electromagnetic radiation. If PO is formed via a solid state or surface reaction route at low temperatures, hydrogen deuterium exchange reactions might come into play with significantly enhanced concentrations of deuterium, thus giving important hints to the astrochemical formation processes. The detection of deuterated species and the investigation of their relative abundance to the respective main isotopologue might allow to model chemical pathways on the formation of complex organic molecules (COMs) in space \citep{Herbst.2009}. \\

\begin{figure}[hbtp]
\centering
\includegraphics[width=\hsize]{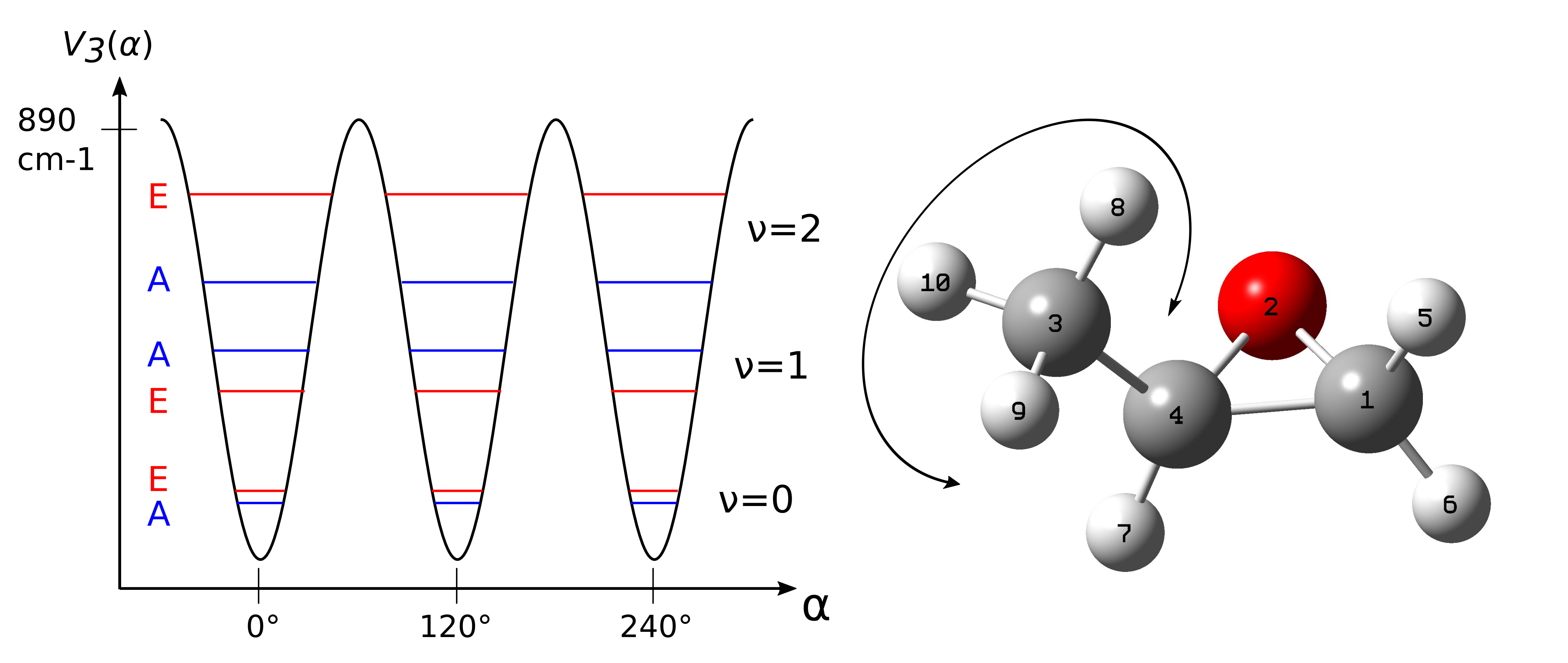}
\caption{Threefold potential function $V_{3}$ for the internal rotation of the methyl group with A and E levels for different excited torsional states (left). The structure of doubly deuterated R-propylene oxide, calculated at the B3LYP/aug-cc-pVTZ level of theory using Gaussian 16\citep{Frisch.2016} (right). The two deuterium atoms are located at positions 5 and 6 at the carbon C-1.}
\label{fig:structure}
\end{figure}

In the presented study we analyze the A-E internal rotational splitting of the doubly deuterated propylene oxide, 
$\mathrm{CH_{3}CHCD_{2}O}$, PO$-$d$_{5,6}$, which has two deuterium atoms substituted to the epoxide ring (see Fig. \ref{fig:structure}). To our knowledge, this is the first study to address the effect of deuterium substitution on the internal torsional motion of the methyl group. The symmetric substitution of the (5,6)-hydrogen atoms maintains the simple structure of a single chirality center. Furthermore, the new results may also foster the search of deuterated PO in cold molecular regions of space. A line list with transition frequencies of $\mathrm{CH_{3}CHCD_{2}O}$ is provided, which helps radio astronomers to search for this molecule in suitable astronomical sources.

\section{Synthesis of the deuterated PO sample} \label{syn}
Racemic PO$-$d$_{5,6}$ was synthesized starting from racemic alanine which was converted to 2-chloropropanoic acid following a published procedure \citep{Koppenhoefer.2003}. 2-Chloropropanoic acid was reduced with \ce{LiAlD4} to 2-chloropropan-1,1-d2-1-ol based on adapted literature procedures \citep{Koppenhoefer.2003b, Karakalos.2016}. Cyclization of neat 2-chloropropan-1,1-d2-1-ol (2.2\,g, 23.3\,mmol) was effected with 2\,g of KOH, dissolved in 5\,ml of water, in a distillation apparatus, where the product, PO$-$d$_{5,6}$, was collected in a trap cooled to -80$\degree$C, at 100\,mbar pressure. To remove water liberated during the reaction, the crude product was treated with \ce{CaH2} (2\,g) at 0$\degree$C until the gas evolution ceased and then was subjected to a second distillation furnishing 1.0\,g (17.2\,mmol, 74\,\%) PO$-$d$_{5,6}$. The product was characterized with ${^1}$H and $^{13}$C-NMR spectroscopy confirming its identity and purity: $\delta$(${^1}$H): 1.31 (d, $^{3}J_{\mathrm{HH}}$ = 5.2\,Hz, \ce{CH3}), 2.97 (q, $^{3}J_{\mathrm{HH}}$ = 5.2\,Hz, CH); $\delta$($^{13}$C{${^1}$H}): 18.1 (s, \ce{CH3}), 47.5 (q, $^{1}J_{\mathrm{CD}}$ = 26.6 Hz, \ce{CD2}), 48.3 (s, CH). 
 
\section{Experimental setup} \label{meas}
Rotational spectra of PO$-$d$_{5,6}$ were recorded from 83$-$110\,GHz, 170$-$220\,GHz, and 260$-$330\,GHz using the Kassel THz spectrometer, which utilises a 2$f$ frequency lock-in modulation technique. The details and data reduction description of this spectrometer are reported elsewhere \citep{Herberth.2019,Stahl.2020}, thus only a brief description of the experiment is given. In our experiment, the deuterated PO sample was probed in a static vacuum glass cell with a total length of 3\,m at an operating pressure of about 1$-$2\,Pa.  The cell was evacuated using a rotary vane pump in combination with a turbo molecular pump. The liquid sample was placed in a long-necked flat-bottomed glass flask, which was connected to the vacuum cell via a needle valve. PO has a high vapour pressure at room temperature\footnote{The main isotopologue has a vapour pressure of about 580\,hPa at room temperature \citep{Bott.1966}}. The needle valve allowed to accurately adjust the operating pressure in the glass cell without heating the sample. In this experiment, it was crucial to operate at a low pressure to resolve the small A-E splitting of the ground state, which otherwise is blended for higher pressures.
 
\section{Results \& Discussion}
\subsection{Quantum chemical calculations}
Anharmonic frequency calculations on the B3LYP/aug-cc-pVTZ level of theory were performed using the computational chemistry program Gaussian 16 \citep{Frisch.2016}.
The rotational constants and the centrifugal distortion constants up to sextic order were determined and are summarised in Table \ref{tab:results1}.
A harmonic hindered rotor calculation of the barrier height to internal rotation resulted in $V_{3}^{harm}=2.4\,\mathrm{kcal}=10.04\,\mathrm{kJ}=839.44\,\mathrm{cm^{-1}}$ (1\,atm, $T=298.15$\,K, 1\,mol). The electric dipole moment components were calculated from the equlibrium structure of PO$-$d$_{5,6}$ and are $\mu_{a}=0.87$\,D, $\mu_{b}=1.72$\,D, $\mu_{c}=0.59$\,D, and the total dipole moment is $\mu=2$\,D.
\subsection{Spectral analysis}
\begin{figure*}[!ht]
\centering
\includegraphics[width=\hsize]{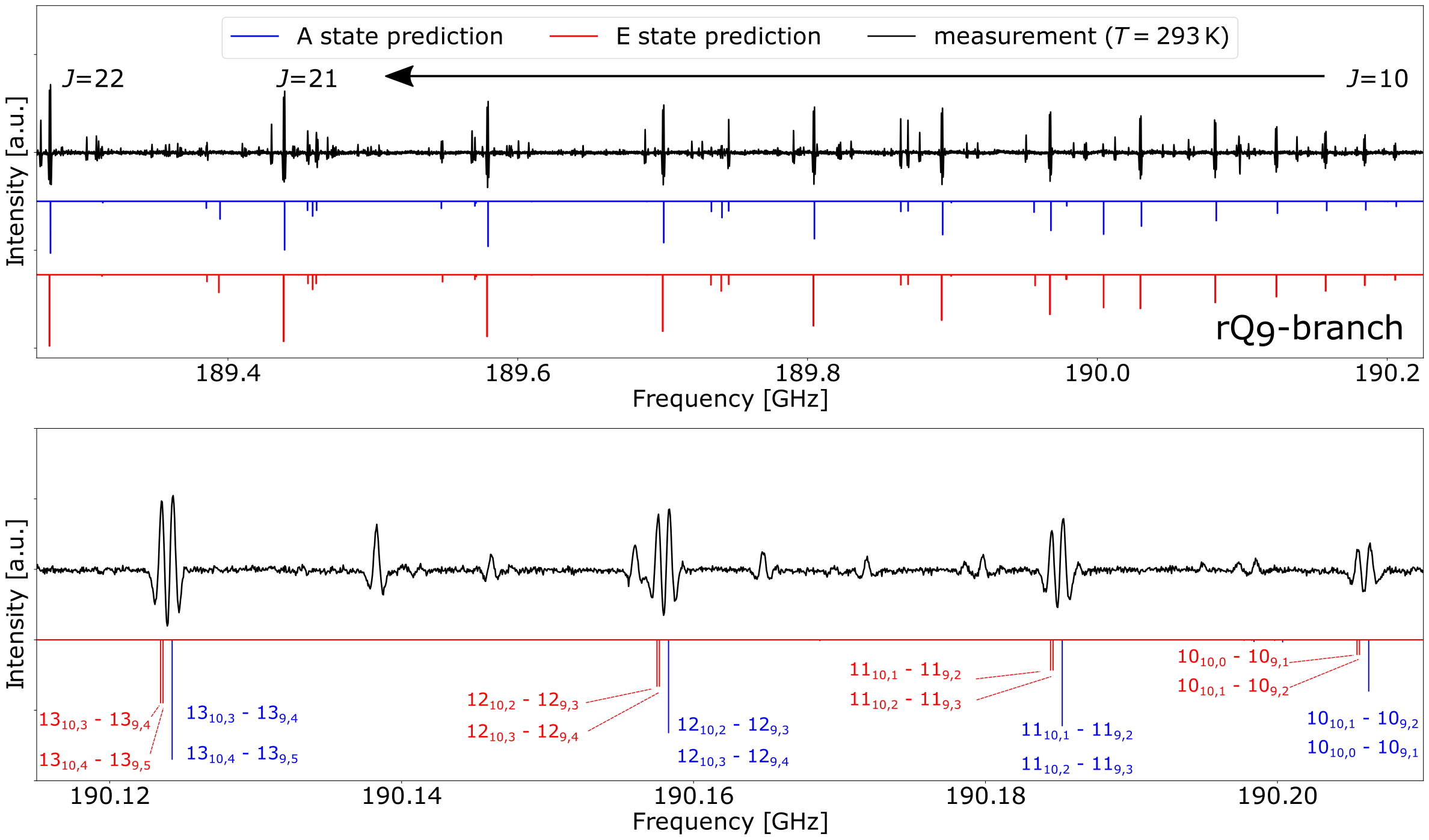}
\caption{
Spectrum of deuterated propylene oxide around 190\,GHz measured in a static cell at room temperature. Top graph: the rQ$_{9}$-branch from $J''=10$ to $J''=22$ can be seen as a measured spectrum (black), and as a stick spectrum prediction of the A (blue) and E (red) states, based on a least-squares-fit analysis with XIAM (Tables \ref{tab:results1} and \ref{tab:results2}). Bottom graph: the detailed view of the rQ$_{9}$-branch shows the splittings of a rotational transitions into A and E components.  
}
\label{fig:Qbranch}
\end{figure*}

\begin{figure*}[!ht]
\centering
\includegraphics[width=\hsize]{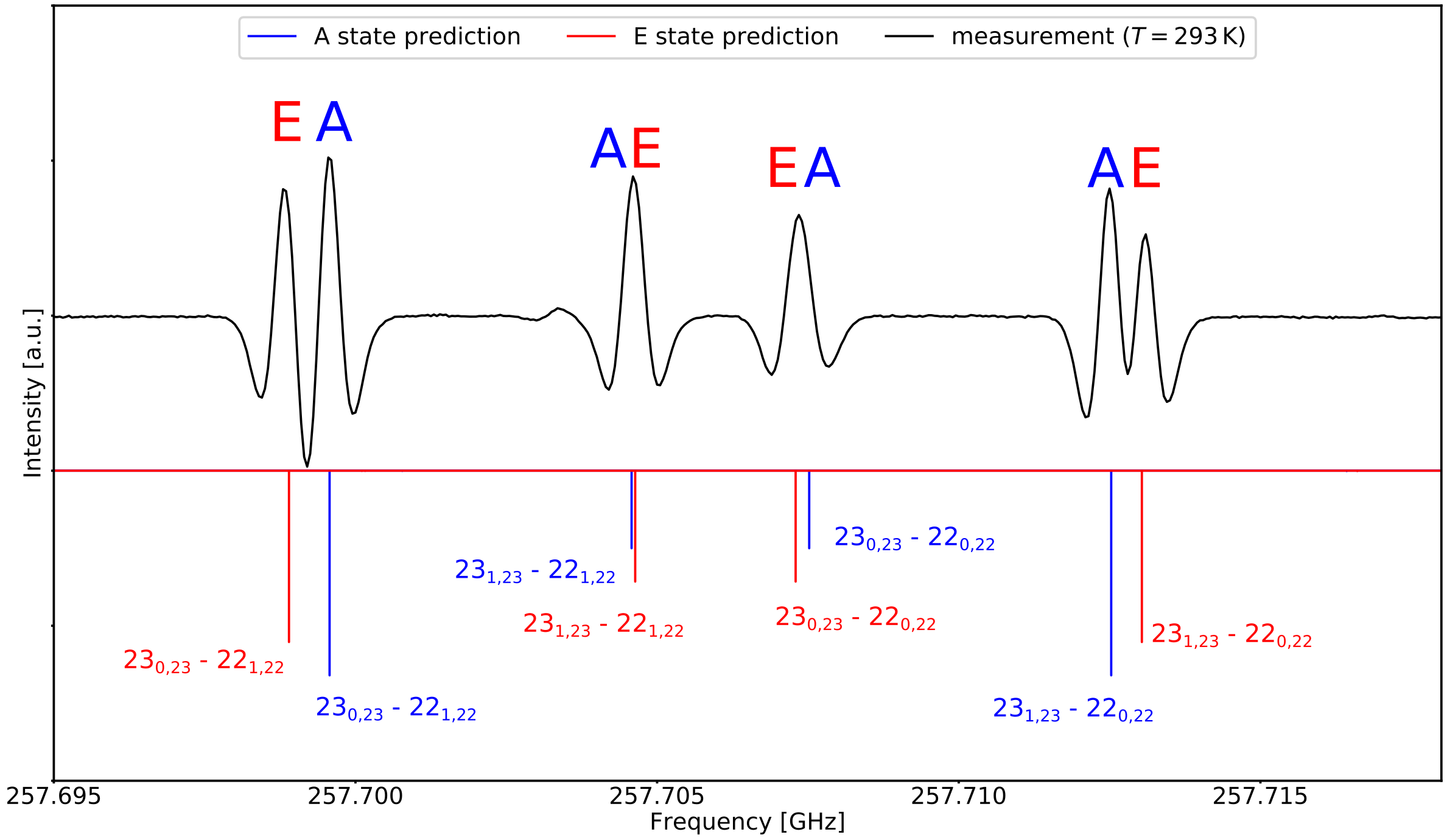}
\caption{
Spectrum of deuterated propylene oxide around 257.7\,GHz measured in a static cell at room temperature. R-branch transitions with $J''=22$ can be seen in the measured spectrum (black) with the stick spectrum prediction of the A (blue) and E (red) states, based on a least-squares-fit analysis with XIAM (Tables \ref{tab:results1} and \ref{tab:results2}). The transitions are labeled by their quantum numbers. The outer pairs of transitions are \textit{b}-type transitions, which are split up into their A and E components. In the case of the inner lines, which are \textit{a}-type transitions, the A and E states are blended.
}
\label{fig:Rbranch}
\end{figure*}

\begin{figure*}[!ht]
\centering
\includegraphics[width=\hsize]{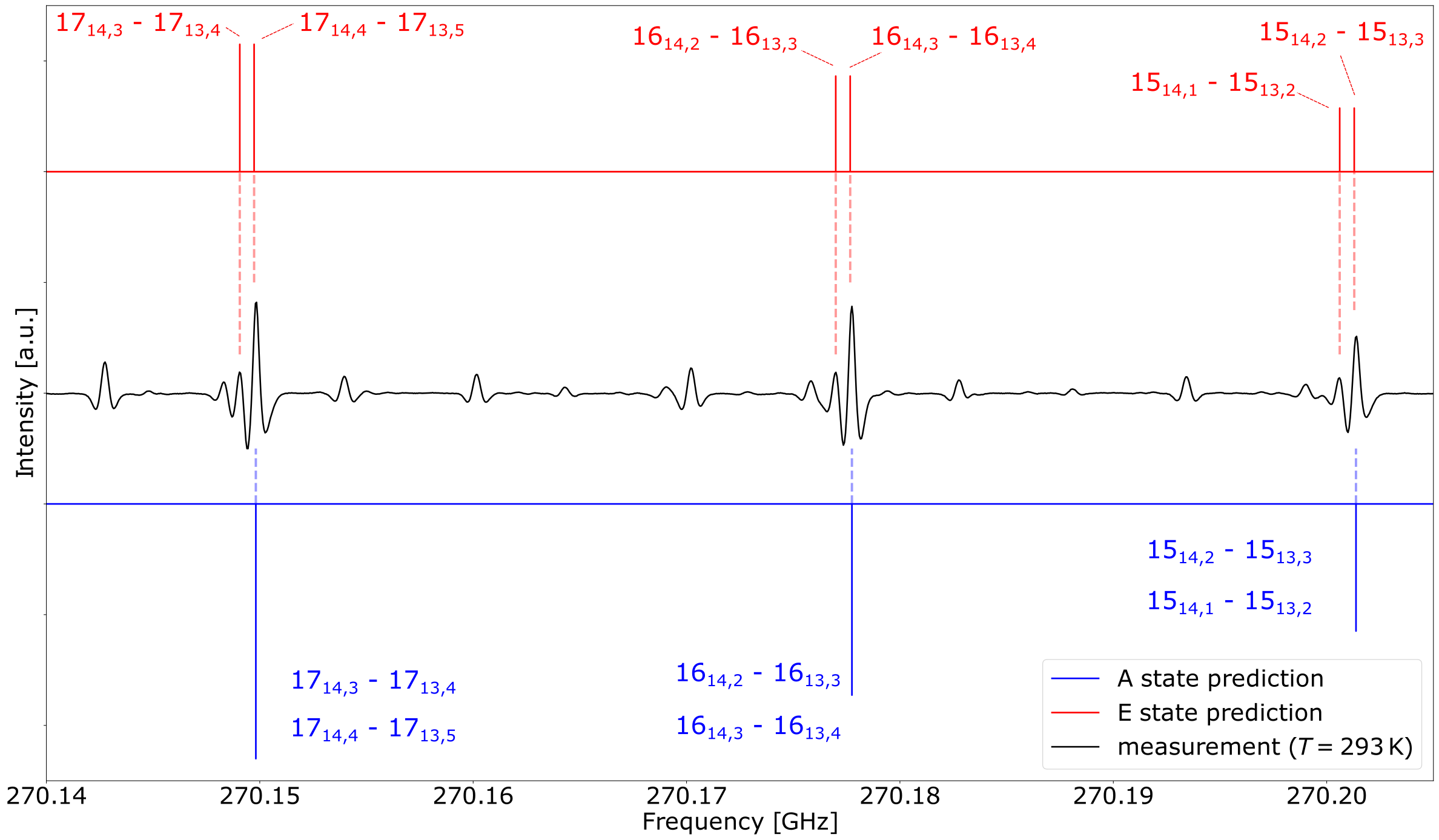}
\caption{
Spectrum of deuterated propylene oxide around 270\,GHz measured in a static cell at room temperature. The rQ$_{13}$-branch transitions from $J''=15$ to $J''=17$ can be seen in the measured spectrum (black) with the stick spectrum prediction of the A (blue) and E (red) states, based on a least-squares-fit analysis with XIAM (Tables \ref{tab:results1} and \ref{tab:results2}). The transitions are labelded by their quantum numbers.
The A state asymmetry sides are blended, whereas the E state asymmetry sides are split, with one of them is blended with the respective A state line.  
}
\label{fig:blended_Qbranch}
\end{figure*}

 At the beginning of the assignment procedure of the A-E splittings of $\mathrm{CH_{3}CHCD_{2}O}$, the strong A state transitions, which resemble those of an asymmetric rigid rotor without internal rotation \citep{Swalen.1957}, were assigned using the software PGOPHER \citep{Western.2014,Western.2017}. For the initial prediction of the A states, the rotational constants and centrifugal distortion constants from the anharmonic frequency calculations were taken. Hundreds of A state transitions were assigned, until a robust fit of the A state was achieved. An initial prediction of both the A and E state transitions was obtained using the program XIAM, where the  barrier height to internal rotation $V_{3}$ and the structural parameters $\rho$, $\beta$, and $\gamma$, which describe the internal rotor axis relative to the remaining frame of the molecule, were included. The initial values of these parameters were based on the knowledge of previous work on the main isotopologue   \citep{Herschbach.1958,Herschbach.1959,Mesko.2017,Stahl.2021}. For further assignments, we then utilised the AABS package \citep{Kisiel.2005, Kisiel.2012} from the PROSPE website \citep{Demaison.2001}, which is able to handle torsional quantum number labels, hence, to assign A and E levels in a combined line list. A and E state transitions were assigned to the line list from the XIAM prediction with the help of AABS. The AABS assignments were fit with XIAM to obtain a more accurate prediction, and the AABS line list was complemented by new assignments from the observations. Iteratively, the XIAM parameters were fit and refined to the successively updated line list until a robust parameter set for the XIAM Hamiltonian was found. The conversion between the XIAM format and the AABS format was achieved with the help of a homemade python code. The final prediction and the experimental results are presented below.
\\
The spectrum of the ground torsional state of PO$-$d$_{5,6}$ shows Q-branches and strong R-branch transitions. At $T=300$\,K, the maximum intensity of the spectrum is between 600$-$800\,GHz with strong R-branch transitions dominating the spectrum above 350\,GHz. The spacing of the Q branches is, to a good approximation, given by $2\cdot(A-\dfrac{B+C}{2})\approx 20\,\mathrm{GHz}$, which is clearly distinct from the 23.4\,GHz spacing of the main isotopologue \citep{Stahl.2021}.
\\
 Rotational levels of low $K_a$ are split by the molecular asymmetry. Due to internal rotation of the methyl group each asymmetry level is further split into one component of non-degenerate A symmetry and one of degenerate E symmetry which leads to A and E rotational transitions of comparable intensity ratios (A:E=1:1) \citep{LIN.1959}. For high $K_a$ states the asymmetry splitting vanishes, resulting into two A components of same energy and two separated components of E symmetry. The spectra of high $K_a$ states show two separated lines (E) and one single line (A) of twice the intensity (E:E:A=1:1:2). This behavior was described in detail by Herschbach \& Swalen in Ref. \citep{Swalen.1957,Herschbach.1958}.  The A-E splitting strongly varies with $J$, $K_a$. For the ground state transitions of PO$-$d$_{5,6}$ the splitting is typically below 1 MHz, leaving some of the splittings unresolved.    
Example spectra and the simulated room temperature spectra based on parameters derived from a least-squares-fit analysis with XIAM (Tables \ref{tab:results1} and \ref{tab:results2}) are shown in Figures \ref{fig:Qbranch}, \ref{fig:Rbranch}, and \ref{fig:blended_Qbranch}.
Figure \ref{fig:Qbranch} shows the rQ$_{9}$-branch with transitions from $J''=10$ to $J''=22$ around 190\,GHz (with $J''$ denoting the quantum number of the lower rotational level).   A stick spectrum prediction of the A and E  states\footnote{In this work, the A levels are marked in blue, the E levels are marked in red.}, based on a least-squares-fit analysis with XIAM (Tables \ref{tab:results1} and \ref{tab:results2}), is also given. The lines are assigned and labeled by asymmetric rotor quantum numbers. Note, that for high $K_a$ states the line splitting is caused by the internal rotation rather than by asymmetry splitting. To be consistent with the assignment of transitions used in our fit routines we kept the labeling of asymmetric top molecules, although the involved energy states are mixtures of both asymmetry components. In the detailed view of the Q-branch in Figure \ref{fig:Qbranch}, the small A-E splitting of the \textit{b}-type transitions with $J''=12$ can be seen. The A states are blended, whereas the predicted E states are split by about 140\,kHz. In the measured spectrum, the doublets of A and E states are equal in  intensity. There are more lines in the observed spectrum than lines predicted by XIAM. These lines could belong to higher torsionally excited states or to impurities in the sample cell.
 
Figure \ref{fig:Rbranch} shows four doublets of A and E state transitions in the R-branch with $23_{K'_{a},23}\leftarrow 22_{K''_{a},22}$ . The outer doublets in Figure \ref{fig:Rbranch} are \textit{b}-type transitions, which are split into well-separated A and E components. The $23_{0,23}\leftarrow 22_{1,22}$ doublet has a splitting of about 800 kHz, the $23_{1,23}\leftarrow 22_{0,22}$ doublet has a splitting of 600 kHz. In contrast, the A and E states of the two inner lines of Figure \ref{fig:Rbranch} are \textit{a}-type transitions with $23_{1,23}\leftarrow 22_{1,22}$, and $23_{0,23}\leftarrow 22_{0,22}$, which A and E components are not resolved, but line broadening due to the A-E splitting of the  $23_{0,23}\leftarrow 22_{0,22}$ transition can be seen. 
In Figure \ref{fig:blended_Qbranch}, Q$_{13}$-branch transitions with $J''=15$, $J''=16$, and $J''=17$ can be seen together with the predicted spectrum from the XIAM program. The asymmetry splitting of A states has vanished, whereas the E state transitions are split by 790\,kHz. As can be seen from the line intensities, one of the E components is blended with the corresponding A component.  
   
\subsection{Spectral results \& discussion}
The results of our investigation of the A-E splitting due to internal rotation with the help of the program XIAM are shown in Tables \ref{tab:results1} and \ref{tab:results2}. 
\begin{table*}[!ht]
\centering
\caption{Rotational constants and centrifugal distortion constants of $\mathrm{CH_{3}CHCD_{2}O}$ from the XIAM analysis. The Hamiltonian is described in Watson's A-reduction and in $I^{r}$ representation. Quantum chemical calculations, and literature values from the main isotopologue are also given for comparison. 
The rotational constants from Imachi \& Kuczowski \citep{Imachi.1983} (originally taken from Reference 8 in \citep{Imachi.1983}) are calculated from the moments of inertia by using the factor 505379.05 MHz/amu$\AA^2$.
}
\begin{tabular}{ccccc}
\hline 
Parameter&  This work (XIAM) & Imachi \& Kuczowski &B3LYP/aug-cc-pVTZ \\
  \hline
$A$ /MHz & \tablenum{15916.944077(329)} & \tablenum{15916.90897} &\tablenum{16076.69}\\ 
$B$ /MHz & \tablenum{6246.890290(572)} &\tablenum{6246.59075 }&\tablenum{6224.51}  \\ 
$C$ /MHz & \tablenum{5545.600172(584)}  &\tablenum{5545.75072}& \tablenum{5528.22}\\ 
$\Delta_{J}$ /kHz   & \tablenum{2.468321(137)}   & & \tablenum{2.44}   \\ 
$\Delta_{JK}$ /kHz  &  \tablenum{3.122962(310)}&&  \tablenum{3.19 } \\ 
$\Delta_{K}$ /kHz  &   \tablenum{14.387857(1143) }& & \tablenum{14.46 }   \\ 
$\delta_{J}$ /kHz   & \tablenum{0.213706(18)}&& \tablenum{0.21} \\ 
$\delta_{K}$ /kHz  &  \tablenum{2.962002(632)} & & \tablenum{2.92}  \\ 
$\Phi_{J}$ /Hz      &  \tablenum{0.000820(38) } & &  \tablenum{ 0.0014} \\ 
$\Phi_{JK}$ /Hz   &							  & & \tablenum{0.0019 } \\ 
$\Phi_{KJ}$ /Hz    &  \tablenum{ -0.008553(754) } & & \tablenum{-0.0087} \\   
$\Phi_{K}$ /Hz  & \tablenum{ 0.214783(2840)  } && \tablenum{  0.0645 } \\ 
$\phi_{J}$ /Hz    &  \tablenum{ 0.000155(4) } & & \tablenum{ 0.0002} \\  
$\phi_{JK}$ /Hz  &	\tablenum{ 0.002964(208) }							 &&\tablenum{ 0.0050 }   \\  
$\phi_{K}$ /Hz  & \tablenum{ 0.100146(1277) } & & \tablenum{ 0.1324 }  \\ 
\hline
No. of lines $N$ & \tablenum{4199} &  &  \\ 
$\sigma$ /kHz  & \tablenum{ 81.7 }&  \multicolumn{2}{c}{$\sqrt{\frac{1}{N}\sum \left(x_{obs}-x_{calc}\right)^2}$} \\ 
\hline
\end{tabular} 
\label{tab:results1}
\end{table*}
\begin{table*}[!ht]
\centering
\caption{Tunneling and structure parameters of $\mathrm{CH_{3}CHCD_{2}O}$ from the XIAM analysis. The Hamiltonian is described in Watson's A-reduction and in $I^{r}$ representation. Literature values from Refs. \citep{Stahl.2021,Mesko.2017,Swalen.1957,Herschbach.1958} for the main isotopologue are also given for comparison. 
}
\begin{tabular}{ccccc}
\hline 
Parameter&  This work (XIAM) & Stahl \textit{et al.}$^{a}$ \citep{Stahl.2021}&Mesko \textit{et al.}$^{a}$ \citep{Mesko.2017} &Herschbach and Swalen$^{a}$ \citep{Swalen.1957,Herschbach.1958} \\
  \hline
$V_{3}$/$\mathrm{cm^{-1}}$  &  882.7835(1.7179) & 898.6611(894) & 892.71 (58)&  895(5) \\ 
$\rho\cdot 10^{-3}$  & 92.248198$^{b}$ & 102.562756& 102.248991 & 103 \\ 
$\beta$ /rad &  -0.1707(11) &  0.1647 &0.1726& \\ 
$\gamma$ /rad  &  -1.7013(351) &1.574 &1.5457& \\ 
$\epsilon$ /rad & 1.4548$^{b}$& 1.574(80) &1.55 (12) & \\ 
$\delta$ /rad & 0.4588$^{b}$ &0.46661(12) &0.4858 (19)&  \\ 
$\angle(i,a)$ /$^\circ$ &  26.2900(1690) &26.7345(65) &27.8368(1098) & 26.8616702 \\ 
$\angle(i,b)$ /$^\circ$ &  87.0610(8049) &90.1(2.1) &89.4020(3.3326) & 89.7364385 \\ 
$\angle(i,c)$ /$^\circ$ &  63.9002(2563) & 63.2656(1052) &62.1708(2602) & 63.1407711 \\ 
$F_{0}$ /GHz  &  156.98027(26010)& 159.0197(1628) &158.2278$^{c}$ & 158.2278  \\ 
$F$ /GHz  &  172.211630665$^{b}$ & 176.281928 &175.277860737 & 175.55846 \\ 
$I_{\alpha}$  & 3.21938(533) & 3.178091(325) &3.193997$^{d}$& 3.194 \\ 
$s$   &  68.301456$^{b}$  & 67.924493 &67.861583$^{d}$ & 68.0 \\ 
\hline
\multicolumn{5}{l}{$^{a}$ These parameters are the values from the main isotopologue $\mathrm{CH_{3}C_{2}H_{3}O}$.} \\
\multicolumn{5}{l}{$^{b}$ These parameters were derived from the fit parameters in the XIAM analysis.} \\
\multicolumn{5}{l}{$^{c}$ The $F_{0}$ value in  Mesko \textit{et al.} was fixed to the value of \citep{Herschbach.1958}.} \\
\multicolumn{5}{l}{$^{d}$ Taken from the XIAM output file} \\
\multicolumn{5}{l}{ ~~ Mesko \textit{et al.} \citep{Mesko.2017} under \url{https://doi.org/10.1016/j.jms.2017.02.003}.}\\
\end{tabular} 
\label{tab:results2}
\end{table*}
The rotational constants $A$, $B$, and $C$, the quartic centrifugal distortion parameters $\Delta_{J}$, $\Delta_{JK}$, $\Delta_{K}$, $\delta_{J}$ and $\delta_{K}$, and the sextic centrifugal distortion constants $\Phi_{J}$, $\Phi_{KJ}$, $\Phi_{K}$, $\phi_{J}$, $\phi_{JK}$, and $\phi_{K}$ were used in Watson's A-reduction in I$^{r}$ representation. Moreover, the barrier height to internal rotation $V_{3}$ and structure parameters have been included to describe the line splitting of A and E components. 
In the XIAM analysis, the A and E states were fitted to an uncertainty level of 82\,kHz, including 4199 transitions with $J,K_{a},K_{a}\leq 67, 28, 62$ between 83\,GHz and 330\,GHz. The obtained fit parameters $V_{3}$, $\beta$, and $\gamma$, and the reduced rotational constant $F_{0}$ allowed to derive the rho axis vector $\rho$, the angles between the principal axis and the internal axis $\angle(i,a/b/c)$, the internal moment of inertia $I_{\alpha}$, and the dimensionless parameter $s$. 
A description of the internal rotor parameters and the XIAM angles can be found in Refs. \citep{LIN.1959,Lister.1978} and \citep{Hartwig.1996,Kleiner.2010}, respectively. 
The additional XIAM tunneling parameters, $\Delta_{\pi2J}$, $\Delta_{\pi2K}$, $\Delta_{\pi2-}$, $D_{c3J}$, $D_{c3K}$, and $D_{c3-}$ \cite{Hansen.1999,Herbers.2020,Herbers.2020b} did not improve the fit and were therefore omitted. This also holds for the $\Delta_{\pi2K}$, which was used for the main isotopologue study \citep{Mesko.2017}.
\\
\\
Imachi \& Kuczowski \citep{Imachi.1983} determined the structure of propylene oxide from an isotopologue analysis, including  three $^{13}$C and several D substitutions. They also reported the moments of inertia\footnote{The moments of inertia reported in Ref. \citep{Imachi.1983} are based on the A state transitions.} of the doubly deuterated propylene oxide with the deuterium atoms at position 5 and 6, which are taken from Ref. 8 in \citep{Imachi.1983}. The rotational constants given in Table \ref{tab:results1} were calculated from the moments of inertia using the factor 505379.05 MHz/amu$\AA^2$. The $A$, $B$, and $C$ reported in Imachi \& Kuczowski are close to the values of this work, for example the relative deviation ($\frac{A-A_{lit}}{A}$, with the literature value $A_{lit}$) for $A$ is $2.2\cdot 10^{-6}$.
   
The rotational constants retrieved from the B3LYP/aug-cc-pVTZ anharmonic frequency calculations have a relative deviation of 1\% and below compared to the experimental results with XIAM. The calculated $A$ value is overestimated by about 160\,MHz, whereas the $B$ and $C$ values are undererstimated by about 20\,MHz compared to the experiment. The quartic centrifugal distortion constants are in good agreement with experiments, but larger deviations are found for the sextic centrifugal distortion constants. The calculated barrier height to internal rotation $V_{3}^{harm}= 839.44\,\mathrm{cm^{-1}}$ is about 40$\,\mathrm{cm^{-1}}$ below the experimental value of $V_{3}=882.7835(1.7179)\,\mathrm{cm^{-1}}$. 
\\
\\
In comparison to the the main isotopologue $\mathrm{CH_{3}C_{2}H_{3}O}$, the barrier height to internal rotation of PO$-$d$_{5,6}$ is lower: Herschbach and Swalen \citep{Herschbach.1958} reported $V_{3}=895(5)\,\mathrm{cm^{-1}}$, Mesko \textit{et al.} \citep{Mesko.2017} reported $V_{3}=892.71(58)\,\mathrm{cm^{-1}}$, which is about 10\,$\mathrm{cm^{-1}}$ higher than for PO$-$d$_{5,6}$. 
The larger uncertainty of 1.7179\,$\mathrm{cm^{-1}}$ in this work compared to 0.58\,$\mathrm{cm^{-1}}$ from Ref. \citep{Mesko.2017} can be explained as follows: the main isotopologue study included 15200 lines up to 1 THz compared to 4199 lines up to 330\,GHz in this work; moreover the value of $F_{0}$ was fixed in Ref. \citep{Mesko.2017}, whereas in this work, the reduced rotational constant was fit, which directly affects the value of $V_{3}$.
Compared to the value $V_{3}=898.6611(894)\,\mathrm{cm^{-1}}$ of the analysis of PO's first excited torsional state \citep{Stahl.2021}, the barrier height of PO$-$d$_{5,6}$ is also lower by about $16\,\mathrm{cm^{-1}}$. Thus, the influence of deuterating  PO's main isotopologue at positions 5 and 6 (see Figure \ref{fig:structure}) on the barrier height to internal rotation is small but not negligible. As was done in the previous work on PO \citep{Herschbach.1957,Herschbach.1958,Mesko.2017,Stahl.2021}, the sixfold potential function $V_{6}$ and higher order contributions to the potential function describing the internal rotation were omitted.  
Deviations from PO$-$d$_{5,6}$ and the main isotopologue \citep{Herschbach.1957,Herschbach.1958,Mesko.2017,Stahl.2021} can also be seen in the other structure parameters, namely, $\rho$, $\beta$, and $\gamma$, and the angles $\angle(i,a/b/c)$. The $\rho$ value of $\cdot 10^{-3}$ is smaller than the main isotopologue values $\rho_{PO}=102.249\cdot 10^{-3}$ \citep{Mesko.2017}, $\rho_{PO}=103\cdot 10^{-3}$ \citep{Swalen.1957,Herschbach.1958}, and $\rho_{PO}=102.52334(449)\cdot 10^{-3}$ and $\rho_{PO}=102.562756\cdot 10^{-3}$ for the ERHAM and XIAM analysis in \citep{Stahl.2021}, respectively. The reduced rotational constants $F_{0}$ and $F$, which are linked to the barrier height to internal rotation, are also smaller than the main isotopologue values. 
Differences of the results of the PO$-$d$_{5,6}$ analysis and its main isotopologue PO can be seen in the comparison of the results with the literature values of the main isotopologue (see Table \ref{tab:results2}). The deuterium atoms affect both the characteristics of the asymmetric rotor, for example the rotational constants have changed, and the internal rotational motion, which can be seen from the structure and tunneling parameters from Tables \ref{tab:results1} and \ref{tab:results2}.
\\
In total, the results of the XIAM analysis of of PO$-$d$_{5,6}$ are in good agreement with the previous work of Imachi \& Kuczowski \citep{Imachi.1983}. The quantum chemical calculations are in good agreement with the experiment. The XIAM prediction, which is based on the least-squares-fit analysis given in Tables \ref{tab:results1} and \ref{tab:results2}, matches the observed spectra and shows an uncertainty of below 82\,kHz. It includes 4199 lines up to 330\,GHz and describes the A-E splitting due to internal rotation. The uncertainty level of our analysis and the inclusion of higher order centrifugal distortion parameters enables the prediction of A and E lines beyond the performed measurements up to 330\,GHz. 
\ \\
The main isotopologue has been found in space towards SgrB2(N) \citep{McGuire.2016}, and a search for deuterated PO species appears promising. Especially, in cold clouds, prestellar and protostellar cores a significantly higher deuterium fractionation compared to the elemental abundance D/H$=(1.5\pm 0.1)\cdot 10^{-5}$ \citep{Linsky.1998,Roueff.2005,Herbst.2009} can be expected. Ethylene oxide, the simplest epoxide $\mathrm{C_{2}H_{4}O}$, has been detected in space towards several sources, among them SgrB2(N)
 \citep{Dickens.1997} 
 and the solar-type protostellar binary IRAS 16293-2422 \citep{Lykke.2017}. The millimeter and submillimeter spectra of its singly deuterated isotopologues are known from literature \citep{Albert.2019}, but yet these species have not been detected in space. Although the methyl radical $\mathrm{CH_{3}}$ was observed in space \citep{Feuchtgruber.2000} it is unlikely that it is directly involved in the propylene oxide formation. Other methyl bearing molecules like methanol $\mathrm{CH_{3}OH}$ are detected towards SgrB2(N) \citep{Ball.1970} and many other sources.
In the case of SgrB2(N) only a tentative detection of the deuterated species $\mathrm{CH_{2}DOH}$ is reported \citep{Belloche.2016}.   
However, methanol can be highly deuterated, like the triply-deuterated methanol found towards IRAS 16293-2422, i.e. in the same source where also ethylene oxide has been detected \citep{Parise.2004}. The spectra around 10$-$40\,GHz of singly deuterated species of propylene oxide have been studied by Imachi \& Kuczkowski \citep{Imachi.1983}. If propylene oxide is found in IRAS 16293-2422 or in similar sources, deuteration might occur more readily and might also allow for doubly deuterated species, like $\mathrm{CH_{3}CHCD_{2}O}$. For a possible detection,
we provide a line list with A and E level transitions of the ground state of PO$-$d$_{5,6}$  in the millimeter and submillimeter-wave range. In case of its astronomical detection, the new measurements will enable the calculation of accurate column densities which further elucidate astrochemical formation processes of chiral molecules in the interstellar medium.     

\section{Conclusions}
In this work, the high-resolution internal rotation analysis of the ground state of the chiral molecule $\mathrm{CH_{3}CHCD_{2}O}$ is presented. The least-squares-fit analysis using XIAM included 4199 A and E state transitions between 83\,GHz and 330\,GHz. The overall Hamiltonian in XIAM with the parameters from Tables \ref{tab:results1} and \ref{tab:results2} describes the A-E splitting up to sub-millimeterwave frequencies with an uncertainty of 81.7\,kHz. Furthermore, the barrier height to internal rotation $V_{3}$ was determined to be $V_{3}=882.7835(1.7179)\,\mathrm{cm^{-1}}$. Anharmonic frequency calculations were performed on the B3LYP/aug-cc-pVTZ level of theory, which are in good agreement with the experiment with rotational constants up to quartic centrifugal distortion parameters; the calculated sextic centrifugal distortion parameters showed larger deviations but they were not relevant in the initial assignment procedure. The rotational constants are in good  agreement with the values reported in Ref. \citep{Imachi.1983}. The results of PO$-$d$_{5,6}$  were compared to those of the main isotopologue, and the influence of the deuterium substitution on the structure and internal rotational motion was briefly discussed.

\section{Acknowledgements}
TFG, RP, DK and PS acknowledge the funding by the Deutsche Forschungsgemeinschaft (DFG, German Research Foundation) – Projektnummer 328961117 – CRC 1319 ELCH.  PS and GWF also gratefully acknowledge the funding from the DFG-FU 715/3-1 project.

\end{document}